\documentclass{svproc}
\usepackage{url}

\usepackage{amssymb}
\usepackage{amsmath}
\usepackage{mathtools}
\usepackage{xcolor}

\begin{document}
\mainmatter              
\title{Archetypal analysis of European 10-year government bond yields with multidimensional scaling of two-mode three-way asymmetric dissimilarities}
\titlerunning{Archetypal analysis of bond yields by asymmetric dissimilarities}  
%
\author{Aleix Alcacer\inst{1} \and Rafael Ben\'itez\inst{2} \and
Vicente J. Bol\'os\inst{2} \and Irene Epifanio\inst{1}\thanks{Corresponding author}}
\authorrunning{Aleix Alcacer et al.} 
%
\tocauthor{Aleix Alcacer, Rafael Ben\'itez, 
Vicente J. Bol\'os and Irene Epifanio}
\institute{Jaume I University, Castell\'o 12071, Spain.\\
\and
University of Val\`{e}ncia,
Val\`{e}ncia 46022, Spain.\\
\email{epifanio@uji.es}}

\maketitle              

\begin{abstract}
A recent methodology for extracting archetypal profiles from three-way asymmetric proximity data is applied to a dataset comprising $23\times 23\times 3$ two-mode, three-way asymmetric dissimilarity matrices. The asymmetric dissimilarities are based on the oriented wavelet squared coherence of 10-year government bond yields among 23 European countries across three time intervals from 2001 to the present. First, the h-plot is computed for the unconditional two-mode three-way data, which is then represented in a unified Euclidean space, providing an intuitive and interpretable visualization. Subsequently, archetypoid analysis is performed. This unsupervised methodology identifies the archetypal countries and expresses all remaining countries as mixtures of these archetypal instances. Additionally, the degree of asymmetry for each country is calculated, offering further insight into the structure of the dissimilarities. The dataset and code are provided to support reproducible research.
\end{abstract}
\section{Introduction}
In recent years, dimensionality reduction techniques for modeling asymmetric relationships have attracted increasing attention, as evidenced by the rising volume of related publications \cite{bove2018methods,bove2021methods,okada2024applied}.

 For three-way asymmetric proximity data, relatively few clustering methods have been proposed \cite{bocci2024clustering,okada2025analysis}, and only one, \cite{alcacer2025multidimensional}, has been introduced recently that addresses both archetypal analysis \cite{cutler1994archetypal} and clustering. In \cite{alcacer2025multidimensional}, the h-plot methodology \cite{Epi2013,Epi2014} is generalized to three-way proximity data, accommodating symmetric and asymmetric structures as well as conditional and unconditional formulations. This framework provides a unified Euclidean representation that enhances interpretability, together with a closed-form, eigen-decomposition–based solution that circumvents local optima. Additional benefits include invariance to linear scaling, computational efficiency for high-dimensional matrices, and an accessible measure of goodness of fit. Moreover, by embedding asymmetric and non-reflexive dissimilarities into a Euclidean space, the h-plot supports the application of complementary methods, such as archetypoid analysis \cite{Vinue15}, for the identification of representative extreme profiles.

This study represents the first application of the methodology proposed in \cite{alcacer2025multidimensional} to macroeconomic sovereign bond data, specifically 10-year government bond yields across 23 European countries over three distinct time intervals from 2001 to the present. This approach is particularly well-suited for the macroeconomic context, as it enables the analysis of three-way asymmetric dissimilarities among countries and provides an interpretable, Euclidean representation of complex temporal relationships in sovereign bond markets. The analysis is based on a dataset consisting of $23 \times 23 \times 3$ two-mode, three-way asymmetric dissimilarity matrices constructed using wavelet squared coherence \cite{Ferrer21}, considering the unconditional framework, i.e. when entries from distinct matrices are directly comparable. Section \ref{met} presents the methodology, Section \ref{res} describes the data and reports the results, and concluding remarks are provided in Section \ref{con}. Code and data are available at {\url{https://epifanio.uji.es/RESEARCH/asyhploteuro.zip}}.

\section{Methodology} \label{met}
\subsection{Multidimensional scaling for unconditional 2-mode 3-way data}
Let $\mathbf{\Delta}_l$ ($l = 1, \ldots, L$) be a square dissimilarity matrix, where each element $\delta_{ijl}$ denotes the pairwise dissimilarity from object $i$ to object $j$ ($i, j = 1, \ldots, n$) observed under condition or occasion $l$. The matrix is asymmetric, i.e. $\delta_{ijl} \neq \delta_{jil}$. Firstly, we define $
\mathbf{D}$ = $\left[ \mathbf{\Delta}_1 \, \middle| \, \mathbf{\Delta}_1^\top \, \middle| \, \dots \, \middle| \, \mathbf{\Delta}_L \, \middle| \, \mathbf{\Delta}_L^\top \right]$, 
as an $n \times 2Ln$ matrix obtained by concatenating the matrices $\mathbf{\Delta}_l$ and their transposes $\mathbf{\Delta}_l^\top$ column-wise.  

Secondly, the matrix $\mathbf{D}$ is h-plotted in two dimensions \cite{CorstenGabriel}, resulting in the matrix $\mathbf{H}_2$ with $2Ln$ rows and two columns. A two-dimensional h-plot is constructed via eigendecomposition of the variance–covariance matrix $\mathbf{S}$ of $\mathbf{D}$. Let $\lambda_1$ and $\lambda_2$ be the two largest eigenvalues with corresponding unit eigenvectors $\mathbf{q}_1$ and $\mathbf{q}_2$. The h-plot is then represented as $
\mathbf{H}_2$ = $\left( \sqrt{\lambda_1}, \mathbf{q}_1,\ \sqrt{\lambda_2}, \mathbf{q}_2 \right)$.

In this reduced two-dimensional space, the Euclidean distance between rows $\mathbf{h}_i$ and $\mathbf{h}_j$ approximates the sample standard deviation of the difference between variables $i$ and $j$ \cite{Epi2013}. In summary, in the h-plot, the dissimilarity matrix is interpreted as a data matrix with different variables: the dissimilarities from element $j$ to all other elements, denoted $d_{j \cdot}$, and the dissimilarities from all other elements to $j$, denoted $d_{\cdot j}$.

Thirdly, by reversing the concatenation, $\mathbf{H}_2$ is partitioned into individual blocks, each containing $n$ rows. Each of the $2L$ blocks is represented in a two-dimensional plot, using $L$ distinct colors to indicate different conditions, and two font styles, bold and italic, to distinguish the $d_{\cdot jl}$ and $d_{j \cdot l}$ profiles, respectively. The resulting matrix, denoted $\mathbf{Y}$, has $n$ rows and $4L$ columns.

\subsection{Archetypal profiles for two-mode three-way data}
\label{sec:arch}

Archetypal profiles are identified by applying archetypoid analysis (ADA) to the matrix $\mathbf{Y}$, although alternative approaches, such as clustering methods, could also be employed as both approaches aim to uncover structure within data. Nevertheless, clustering and ADA differ conceptually: while clustering summarizes groups using central trends, ADA captures patterns through extreme yet representative instances, as contrasts are often easier to understand. Consequently, clustering is most effective with clearly separated groups, whereas ADA provides more insight when boundaries are diffuse or data form a continuum, as it highlights the extremes defining the variation. Furthermore, ADA yields mixture weights for each subject, detailing its relationship to the archetypoids and offering a nuanced depiction of individual profiles. A comparison of these methods is illustrated by \cite{IsmaelTFM}, demonstrating their respective interpretative advantages.
 
Let us review ADA. ADA models data as convex combinations of a small set of  archetypoids, which correspond to representative actual cases in the data set \cite{alcacer2025survey}. Formally, let 
${\bf Y}$ be an $n \times 4L$ data matrix. ADA introduces three key components:   a) Archetypoids: The $k \times 4L$ matrix $\mathbf{Z}$ contains $k$ archetypoids $\mathbf{z}_j$, each corresponding to an actual instance in $\mathbf{Y}$;  
b) Mixture coefficients: The $n \times k$ matrix $\boldsymbol{\alpha}$ expresses each instance as a convex combination of archetypoids:  
    $\mathbf{y}_i \approx \hat{\mathbf{y}}_i$ = $\sum_{j=1}^k \alpha_{ij} \mathbf{z}_j$, $\alpha_{ij} \geq 0$, $\sum_{j=1}^k \alpha_{ij} = 1$; c) Selection matrix: The $k \times n$ binary matrix $\boldsymbol{\beta}$ indicates which instances are chosen as archetypoids:   
    $\mathbf{z}_j = \sum_{l=1}^n \beta_{jl} \mathbf{x}_l$, $\beta_{jl} \in \{0,1\}$, $\sum_{l=1}^n \beta_{jl} = 1$.

The parameters $\boldsymbol{\alpha}$ and $\boldsymbol{\beta}$ are estimated by minimizing the residual sum of squares, $
RSS$ = $\sum_{i=1}^n \left( \mathbf{x}_i - \sum_{j=1}^k \alpha_{ij} \sum_{l=1}^n \beta_{jl} \mathbf{x}_l \right)^2$, 
subject to the convexity and binary assignment constraints, yielding a mixed-integer optimization problem, which can be solved using the ADA algorithm proposed by \cite{Vinue15} and scaled to larger datasets following the approach in \cite{Vinue21}. For our implementation, we use the R code developed by \cite{EpiIbSi17}. The number of archetypoids is selected using the elbow criterion \cite{Vinue15}, which identifies the point where the RSS shows a marked change in its rate of decrease.

\section{Application} \label{res}

We use a data set corresponding to monthly time series of 10-year government bond yields of 23 European countries obtained from Eurostat: AT (Austria), BE (Belgium), CY (Cyprus), CZ (Czech Republic), DE (Germany), DK (Denmark), EL (Greece), ES (Spain), FI (Finland), FR (France), HU (Hungary), IE (Ireland), IT (Italy), LT (Lithuania), LU (Luxembourg), LV (Latvia), MT (Malta), NL (Netherlands), PL (Poland), PT (Portugal), SE (Sweden), SK (Slovakia), UK (United Kingdom). The time interval runs from January 2001 to April 2025 because many countries do not have data prior to 2001 and, on the other hand, the United Kingdom does not have data available after April 2025 at the time of this analysis. Other countries such as Bulgaria, Estonia and Romania have not been considered because some data for this period is missing. The time interval is divided into three subintervals, taking April 2011 (when the effects of the European debt crisis became apparent \cite{alcacer2026biaafda}) and March 2020 (the start of the COVID-19 crisis) as reference points.

Asymmetric dissimilarities (bounded between 0 and 1) are computed for each time subinterval using the oriented wavelet squared coherence proposed by \cite{Ferrer21}. This measure builds on wavelet squared coherence and incorporates the wavelet phase difference \cite{Torrence99}. The resulting dissimilarities are defined in a causal framework: when ``$i$'' influences ``$j$'', the dissimilarity from ``$i$'' to ``$j$'' is lower than that from ``$j$'' to ``$i$''.

\subsection{Results}

\begin{figure}[t]
\centering
\includegraphics[width=.85\textwidth]{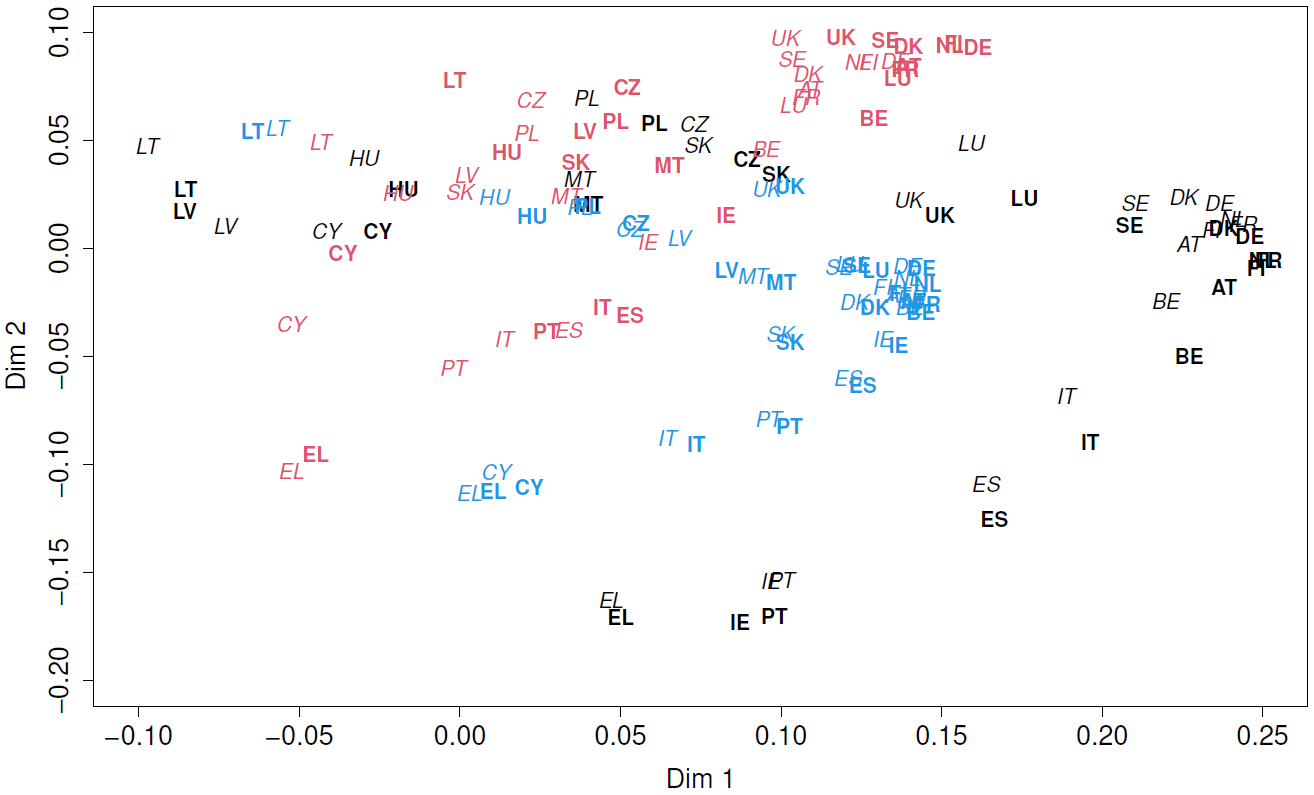}
\caption{H-plot projection of European countries. Bold font represents influenced ($d_{\cdot j}$) profiles, while italic font denotes influencing ($d_{j \cdot}$) profiles. The first time subinterval (from January 2001 to March 2011) is represented in black, the second (from April 2011 to February 2020) in red, and the third (from March 2020 to April 2025) in blue.}
\label{fig:hplot}
\end{figure}

We have computed the h-plot (see Figure \ref{fig:hplot}) for the unconditional two-mode three-way data consisting of three asymmetric dissimilarity matrices (each of which belonging to a different time subinterval) of 23 observations.

Archetypoids are calculated according to Section \ref{sec:arch}. The screeplot shows that the best $k$ is $3$, resulting in the following archetypoids: EL (Greece), HU (Hungary) and NL (Netherlands). The other countries are approximated by a mixture of theses archetypoids, as it is shown in Table \ref{tab:arch}. 

\begin{table}[t]
\centering
\caption{Parameters $\boldsymbol{\alpha}$ multiplied by $100$. This shows the percentage of each archetypoid (1: red, 2: green, or 3: blue) in each country approximation from 2001 to 2025.}
\label{tab:arch}
{\scriptsize
\begin{tabular}{c|rrrrrrrrrrrrrrrrrrrrrrr}
  & \textcolor[rgb]{0.066,0,1}{AT} & \textcolor[rgb]{0.173,0,1}{BE} & \textcolor[rgb]{0.573,1,0}{CY} & \textcolor[rgb]{0,1,0.554}{CZ} & \textcolor[rgb]{0,0.003,1}{DE} & \textcolor[rgb]{0.002,0.075,1}{DK} & \textcolor[rgb]{1,0,0}{EL} & \textcolor[rgb]{1,0,0.962}{ES} & \textcolor[rgb]{0.015,0.002,1}{FI} & \textcolor[rgb]{0.037,0,1}{FR} & \textcolor[rgb]{0,1,0}{HU} & \textcolor[rgb]{1,0,0.839}{IE} & \textcolor[rgb]{1,0,0.944}{IT} & \textcolor[rgb]{0,1,0}{LT} & \textcolor[rgb]{0,0.327,1}{LU} & \textcolor[rgb]{0.003,1,0}{LV} & \textcolor[rgb]{0.061,1,0.466}{MT} & \textcolor[rgb]{0,0,1}{NL} & \textcolor[rgb]{0,1,0.331}{PL} & \textcolor[rgb]{1,0,0.404}{PT} & \textcolor[rgb]{0,0.162,1}{SE} & \textcolor[rgb]{0.168,1,0.732}{SK} & \textcolor[rgb]{0,0.507,1}{UK} \\
\hline
\textcolor[rgb]{1,0,0}{1} & 6                             & 15                            & 36                            & 0                             & 0                             & 1                             & 100                           & 51                            & 2                             & 4                             & 0                             & 54                            & 51                            & 0                             & 0                             & 1                             & 4                             & 0                             & 0                             & 71                            & 0                             & 9                             & 0                             \\
\textcolor[rgb]{0,1,0}{2} & 0                             & 0                             & 64                            & 64                            & 1                             & 7                             & 0                             & 0                             & 1                             & 0                             & 100                           & 0                             & 0                             & 100                           & 25                            & 99                            & 65                            & 0                             & 75                            & 0                             & 14                            & 52                            & 34                            \\
\textcolor[rgb]{0,0,1}{3} & 94                            & 85                            & 0                             & 36                            & 99                            & 92                            & 0                             & 49                            & 97                            & 96                            & 0                             & 46                            & 49                            & 0                             & 75                            & 0                             & 31                            & 100                           & 25                            & 29                            & 86                            & 39                            & 66                           
\end{tabular}
}
\end{table}

The structure of the h-plot is preserved across the three time sub-intervals, exhibiting only some shifts. Greece (archetypoid 1), together with other Southern European countries such as Portugal, Spain, and Italy, occupies a branch located at the lower part of the structure. Countries with a predominance of archetypoid 3 (Netherlands, Germany, Finland, $\ldots$), form a central group. Finally, Baltic and Central European countries, such as Hungary, Poland and the Czech Republic, form another branch at the top characterized by a major weight of archetypoid 2.
Some temporal movements can nonetheless be observed. Ireland is positioned within the Greek branch during the first period but transitions to the central group in the subsequent two periods. Spain also moves closer to this central group during these later periods. Conversely, Cyprus, which is located in the upper branch in the first period, shifts to a position adjacent to Greece by the third period.

With respect to asymmetry (see \cite{alcacer2025multidimensional} for how to assess it from h-plot), the second period exhibits the highest degree of asymmetry, suggesting a greater prevalence of cause–effect relationships during this interval. By contrast, the third period is the least asymmetric. At the country level, Greece displays the lowest degree of asymmetry, which may be interpreted as a more independent evolution of its time series. In contrast, the Baltic countries, together with Luxembourg, show the highest asymmetry, indicating stronger cause–effect relationships in the evolution of their series.

\section{Conclusions} \label{con}

We apply a recently developed extension of the h-plot methodology to the analysis of 10-year government bond yields for a set of European countries over the course of this century. The results identify three country archetypoids: Greece, Hungary, and the Netherlands. All other countries can be represented as mixtures of these archetypoids. Spain, Ireland, Italy, and Portugal are dominated by archetypoid 1; Cyprus, the Czech Republic, Latvia, Lithuania, Malta, Poland, and Slovakia by archetypoid 2; and Austria, Belgium, Germany, Denmark, Finland, France, Luxembourg, Sweden, and the United Kingdom by archetypoid 3.
The extended h-plot provides a unified Euclidean representation with intuitive interpretation, an explicit eigenvector-based solution that avoids local minima, scale invariance under linear transformations, and high computational efficiency for large data sets. In addition, it enables the identification of archetypal profiles and clustering structures, thereby enhancing interpretability and analytical flexibility. Data and code are made publicly available to ensure full reproducibility and transparency.
Several directions for future research arise from this work \cite{alcacer2025multidimensional}. In applied terms, the proposed methodology offers promising potential for empirical research in economics and finance.


\section*{Acknowledgement}
This work was partially supported by the Spanish Ministry of
Science and Innovation PID2023-153128NB-I00, PID2022-141699NB-I00 and PID2020-118763GA-I00 and Generalitat Valenciana
CIPROM/2023/66.

 \bibliographystyle{spmpsci} 
  \bibliography{hplot.bib}

\end{document}